# Optical design of diffraction-limited X-ray telescopes

Brandon D. Chalifoux, Ralf K. Heilmann, Herman L. Marshall and Mark L. Schattenburg*

*Kavli Institute for Astrophysics and Space Research, Massachusetts Institute of Technology,
77 Massachusetts Avenue, Cambridge, MA 02139, USA*
*marks@space.mit.edu

**Abstract:** Astronomical imaging with micro-arcsecond (µas) angular resolution could enable breakthrough scientific discoveries. Previously-proposed µas X-ray imager designs have been interferometers with limited effective collecting area. Here we describe X-ray telescopes achieving diffraction-limited performance over a wide energy band with large effective area, employing a nested-shell architecture with grazing-incidence mirrors, while matching the optical path lengths between all shells. We present two compact nested-shell Wolter Type 2 grazing-incidence telescope designs for diffraction-limited X-ray imaging: a *micro-arcsecond telescope* design with 14 µas angular resolution and 2.9 m² of effective area at 5 keV photon energy (λ=0.25 nm), and a smaller *milli-arcsecond telescope* design with 525 µas resolution and 645 cm² effective area at 1 keV (λ=1.24 nm). We describe how to match the optical path lengths between all shells in a compact mirror assembly, and investigate chromatic and off-axis aberrations. Chromatic aberration results from total external reflection off of mirror surfaces, and we greatly mitigate its effects by slightly adjusting the path lengths in each mirror shell. The mirror surface height error and alignment requirements for diffraction-limited performance are challenging but arguably achieveable in the coming decades. Since the focal ratio for a diffraction-limited X-ray telescope is extremely large ($f/D \sim 10^5$), the only important off-axis aberration is curvature of field, so a 1 arcsecond field of view is feasible with a flat detector. The detector must fly in formation with the mirror assembly, but relative positioning tolerances are on the order of 1 m over a distance of some tens to hundreds of kilometers. While there are many challenges to achieving diffraction-limited X-ray imaging, we did not find any fundamental barriers.

## 1. Introduction

Astronomical imaging with extremely high angular resolution, in the milli-arcsecond (mas) to micro-arcsecond (µas) range, could greatly enhance our understanding of the universe. The Event Horizon Telescope, an Earth-sized interferometer observing at a wavelength of λ=1.3 mm, recently imaged the shadow due to the event horizon of the supermassive black hole in the active galaxy M87 with angular resolution around 25 µas [1]. Imaging with such high resolution in other bands would be valuable, but has not yet been achieved. High-resolution X-ray imaging, in particular, is expected to enable significant advances in astronomy.

With 0.5 arcsecond imaging resolution, the Chandra X-ray Observatory has been an extremely valuable resource for astronomy for the past 20 years [2]. Some examples of the scientific return from Chandra's imaging capability include resolving nearly all of the cosmic X-ray background into point sources (e.g., [3]), finding microlensing due to stars in gravitationally-lensed active galaxies to examine the dark matter fraction in the intervening elliptical galaxy [4], and observing jets from galactic X-ray binaries [5] and high redshift quasars [6]. At higher resolution, we can expect to continue studies such as these; in particular, quasar jets show structure on all scales down to the event horizon in the radio band [7].

At an imaging resolution of order milli-arcseconds, for example, dark matter clumping that is expected based on many cosmological simulations can be tested directly by examining the "speckles" produced in gravitational lenses and, simultaneously, the stellar masses of the lensing galaxies can be measured due to the extremely compact sizes of the X-ray emission regions of active galactic nuclei (AGN). Currently, X-ray light curves are used for these studies (e.g., [8]) but direct imaging of the speckles will provide the distribution of the masses of dark matter sub-halos needed for testing models of the evolution of cosmic structure. At the µas scale, the lensing due to individual stars will discern the mass fraction in stars [9]. There are a handful of gravitationally lensed AGN with X-ray fluxes > 5 x 10⁻¹³

erg/cm$^2$/s [10], which would provide >100 counts in each of 100 speckles in $10^5$ s for an instrument with an effective area of 600 cm$^2$.

The structure of AGN X-ray emission regions themselves is in doubt. There is emission from the corona, which may be either a hot region above the accretion disk that should be approximately spherically symmetric, or a relativistic jet that is aligned with the black hole spin axis. Both variability and reverberation mapping show that the X-ray emission regions are less than 100 times the size of the event horizon [11,12]. To resolve the X-ray emission regions requires angular resolution of order µas but may be the only way to settle this question of geometry and physical origin. There are a few hundred AGN with X-ray fluxes > $10^{-11}$ erg/cm$^2$/s [13] that would yield >100 counts in each of 100 resolution elements in observations of $10^4$ s for an instrument with an effective area of 300 cm$^2$, allowing for variability studies on a Keplerian time scale. Many additional scientific objectives of a µas X-ray imager have been suggested [14].

For µas imaging, space-based interferometry has been proposed in the infrared, visible and ultraviolet bands with kilometer-baselines [15,16] and in the X-ray band with meter-baselines [14,17]. Refractive and diffractive approaches have also been proposed for high-resolution X-ray astronomy [18,19]. In this paper we present optical designs for diffraction-limited reflective X-ray telescopes that employ a nested-shell architecture and can achieve large effective area with mas to µas angular resolution in a compact optical assembly.

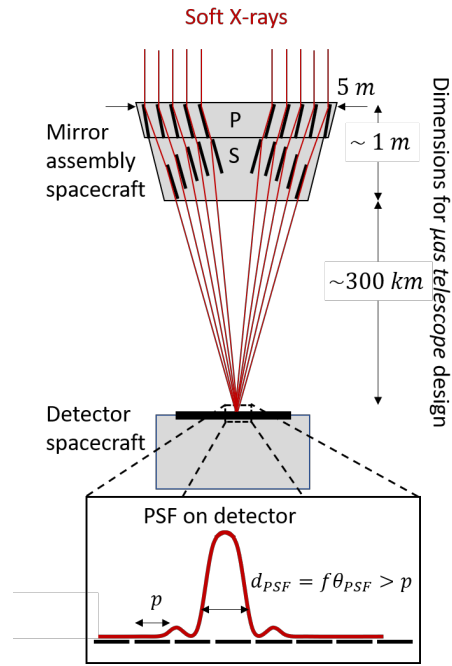

Figure 1. A diffraction-limited nested-shell Wolter Type 2 grazing-incidence X-ray telescope, with approximate dimensions shown for the *µas telescope* design. The mirror assembly and detector are on separate spacecraft flying in formation some hundreds of kilometers apart. The mirror assembly contains curved primary (P) and secondary (S) mirror shells. The inset of the detector spacecraft illustrates that the point-spread function (PSF) should be similar in size to the detector spatial resolution $p$.

X-rays only efficiently reflect off of most surfaces near grazing-incidence, so X-ray telescopes typically use nested grazing-incidence mirror shells (Fig. 1) to achieve large light-gathering power [20,21]. Diffraction-limited imaging or interferometry requires matching optical path lengths throughout the telescope to a fraction of a wavelength, which in turn requires appropriate optical design as well as accurate mirror surfaces and alignment. All previous X-ray interferometer concepts [14,17,19] have been limited in light-gathering power because the optical design led to path lengths that were only matched for mirrors contained in a single shell. With a single shell, the telescope aperture is a very thin annulus and the point-spread function (PSF) has diffraction rings with significant optical power [22]. Matching path lengths of different shells in an X-ray imaging system has been described qualitatively [17], but quantitative methods of specifying the geometric parameters of nested-shell telescopes have not been presented.

In this work, we describe a methodology for designing diffraction-limited nested-shell grazing-incidence X-ray telescopes. Furthermore, we estimate the imaging characteristics of such telescopes, including resolution, effective area, chromatic aberration, and off-axis aberrations. For this analysis, we assume perfect mirror surfaces, but briefly analyze wavefront errors resulting from surface and alignment errors. We also consider pointing knowledge and control requirements. We present two representative Wolter Type 2 optical designs for diffraction-limited X-ray telescopes, a *µas telescope* design (Fig. 1) with 14 µas angular resolution and 2.9 m² of effective area at 5 keV photon energy ($\lambda$=0.25 nm), and a smaller *mas telescope* design with 525 µas resolution and 645 cm² effective area at 1 keV ($\lambda$=1.24 nm). Both designs feature wide diffraction-limited fields of view, large depths-of-focus, and chromatic aberration reduced below the diffraction-limit up to 10 keV.

Fabricating, aligning, and mounting mirrors to achieve diffraction-limited imaging at sub-nanometer wavelengths is a considerable challenge. However, the tightest surface accuracy and alignment requirements for diffraction-limited soft X-ray mirrors are of order 1 nm, which is larger than the allowable path length errors (which must be a small fraction of the wavelength) due to small graze angles [23]. The surface accuracy of thin silicon X-ray mirrors is already approaching this level [24], and there are examples of non-X-ray-telescope optics exceeding this accuracy [25,26].

There are a number of challenges to µas imaging in general, and diffraction-limited X-ray imaging in particular. In Section 2, we briefly discuss the long focal lengths required for µas imaging, then we describe a quantitative design process for diffraction-limited nested-shell X-ray telescopes, and we present the geometry of the two representative designs. In Section 3, we estimate the optical performance of the *µas* and *mas telescope* designs. In Section 4, we address the important issue of chromatic aberration, which arises because of reflection at angles below the critical angle for total external reflection, and which we mitigate by slightly adjusting the path lengths for each shell. In Section 5, we consider off-axis aberrations and related issues of pointing knowledge and control.

## 2. Telescope design

### 2.1 General considerations

Micro-arcsecond imaging, regardless of wavelength, requires extremely long focal lengths in order to realize a practical focal spot size $d_{PSF} = f\theta_{PSF}$, where $\theta_{PSF}$ is the angular diameter of the point spread function (PSF, i.e., the angular resolution, see Section 3) and $f$ is the focal length of the telescope. Furthermore, a diffraction-limited telescope requires a focal ratio of approximately $f/D > p/\lambda$, where $p$ is the detector spatial resolution, $\lambda$ is the photon wavelength, and $D$ is the telescope aperture diameter (this assumes $d_{PSF} \sim \lambda f/D > p$). Diffraction-limited X-ray imaging at $\lambda$=0.25 nm and $p$=25 µm requires $f/D \sim 10^5$, and achieving roughly 20 µas angular resolution requires $f \sim 260$ km and $D \sim 3$ m. The mirror assembly can be compact and launched as a pre-assembled structure, in contrast to longer-wavelength (e.g., visible-band) µas telescope designs that would require much larger-diameter structures. However, as with any µas telescope, the focal length must be long, likely requiring two spacecraft flying in formation hundreds of kilometers apart, one containing the mirror assembly and the other containing the detector (Fig. 1).

The positive and negative implications of using a detector with higher spatial resolution, which would enable a smaller focal length, are discussed in Section 5. Unless the mirror assembly is axially-stretched over tens of meters (as in [14]), or huge improvements in detector spatial resolution (by 4-5 orders of magnitude compared to the state-of-the-art) can be realized, formation flying will be necessary. The GRACE-FO mission has demonstrated formation flying over similar length scales [27], and we show in Section 5 that the required relative position control tolerances of the two spacecraft for an X-ray telescope are on the order of a meter.

Micro-arcsecond imaging at any wavelength requires knowledge of the telescope pointing to µas tolerances, but wavelengths longer than X-rays (e.g., visible) may require pointing control with µas tolerances as well. The relative position and orientation requirements of the two spacecraft are less stringent for an X-ray telescope, for two reasons. First, the extremely large focal ratio results in minimal geometric aberration effects and a large depth of focus. Second, the X-ray photons have sufficiently high energy and low flux to be individually counted and time-tagged. Images can then be re-constructed from the photon counts as long as the pointing *knowledge* is better than the angular resolution. Therefore, X-ray telescopes require µas pointing knowledge but much looser control, as discussed in Section 5. High-precision astrometry to better than 50 µas has been achieved by Gaia [28], and Gravity Probe B has demonstrated pointing knowledge drift on the order of 100 µas/year using a telescope with only a 144 mm diameter aperture [29]. Other potential solutions to pointing knowledge have been proposed [17].

X-rays efficiently reflect off of most surfaces only near grazing incidence, and there are four types of grazing-incidence telescopes [30,31], and Saha [32] presented a useful comparison of these. These telescopes use two subsequent grazing-incidence mirror shells, called the primary and secondary mirrors, that are each nearly-conical. As each shell has a small geometric collecting area defined by a thin annulus, multiple confocal shells are typically nested to increase effective area and thereby improve telescope sensitivity. Achieving diffraction-limited performance requires matching the optical path lengths for all shells to within a fraction of a wavelength. Most existing X-ray telescopes are of the Type 1 design (e.g., [33–35]), while proposed X-ray interferometers have been akin to the Type 2 [14,17] or Type 4 [36] designs but using flat mirror segments to approximate cones. We present a Type 2 telescope (illustrated in Fig. 2) that uses nested shells comprising curved mirrors.

The commonly-used Type 1 design employs two concave mirror shells, and is impractical for µas imaging. Even if the path lengths for all shells in a Type 1 telescope were matched, in order for all shells to be confocal, the graze angle $\alpha$ would need to approximately equal $D/8f$. For a µas-imaging Type 1 X-ray telescope where $f/D \sim 10^5$, the graze angles would need to be of order 1 µrad, and each mirror shell would only contribute a micron-wide annulus. In contrast to the Type 1 design, the Type 2, 3 and 4 designs allow a long focal length that is largely independent of graze angle. The Type 2 telescope is more compact than Types 3 or 4 [32], so we believe a Type 2 telescope is ideal for a µas-imaging telescope. Next, we analyze the geometric parameters of nested-shell Type 2 telescopes in which the path lengths for all shells are equal.

## 2.2  Type 2 telescope optical design

In this section we determine the geometric parameters for nested shells that result in a diffraction-limited telescope, in which all rays passing through the telescope to the focus have the same path length, in accordance with Fermat's principle. Six geometric parameters (Fig. 2) define the geometry of a shell: the focal length $f$, the radius of the leading edge of the primary mirror $r_p$, the radius of the leading edge of the secondary mirror $r_s$, the axial gap between the leading edges of the primary and secondary mirrors $\Delta z$, the telescope length $L$, and the graze angle at the primary mirror $\alpha$. In principle, all of these parameters could vary by shell, but we keep $f$ constant so that the shells are confocal. In Section 2.4, we present a design in which $L$ is constant for all shells. For each shell, one parameter must be chosen as an independent variable, and we use $r_p$ as a convenient choice.

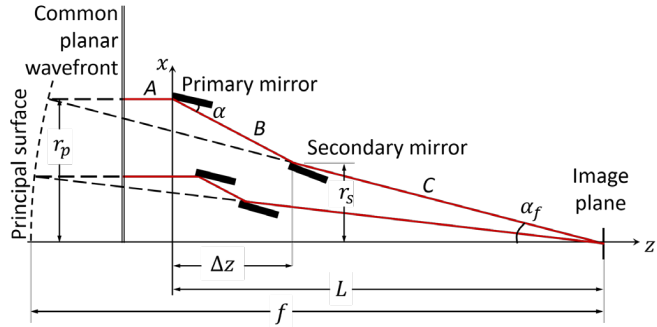

Figure 2. The cross-section of two shells of a Type 2 grazing-incidence telescope (each shell comprises a primary and a secondary mirror), with the geometric parameters labeled for the outer shell only. The mirrors are axisymmetric about the z-axis. Matching path lengths between shells requires matching the length $\mathcal{P} = A + B + C$, which is referenced to a common planar wavefront.

For the telescope to be diffraction-limited, rays through all shells must have the same path length from a common planar wave front (Fig. 2). The length $s = B + C$ can be considered as the length of a straight line (not shown in Fig. 2) connecting the leading edge of the primary mirror to the focus, plus a delay resulting from a jog in this line. The length of this line *increases* with shell radius, so matching the path lengths for all shells requires adding a delay that *decreases* with shell radius. This delay can be visualized as a jog in a line drawn from the primary mirror to the focal plane. The common planar wavefront can be anywhere as long as it is common to all shells. The path length to be equalized is $\mathcal{P} = A + B + C$ as drawn in Fig. 2, where $A$ is the distance between the common planar wavefront and the entrance aperture of a shell, and $s = B + C$ is the path length from the entrance aperture of that shell to the focal plane. Since the focal plane is common to all shells, we choose $A = -L$ for each shell, and thus we must make the quantity $\mathcal{P} = s - L$ equal for all shells in order to achieve diffraction-limited resolution.

Following Saha's derivation of the generalized surface equations and the path length $s$ [32], here we establish three equations that must be satisfied by each shell to match all path lengths $\mathcal{P}$, using the law of reflection, Fermat's principle, the Abbe-Saha sine condition, and an on-axis ray traced from the entrance aperture to the focal plane. The Abbe-Saha sine condition is the Abbe sine condition generalized to account for different optical prescriptions using the parameter $N$, including parabola-hyperbola (PH, $N = 0$), Wolter-Schwarzschild (WS, $N = 1$), and higher-order prescriptions ($N > 1$). Once the parameters for a shell are determined, the surface equations can be calculated as in [32]. The Abbe-Saha sine condition is

$$r_p = \frac{2f \tan(\alpha_f/2)}{1 + N \tan^2(\alpha_f/2)}, \tag{1}$$

where $\alpha_f$ is the angle that a ray makes with the optical axis after reflecting off of the leading edge of the secondary mirror, as shown in Fig. 2. Tracing a ray through the system while using the sine condition – which can be re-arranged as $\tan(\alpha_f/2) = r_p/2f'$, where $2f' = f + (f^2 - Nr_p^2)^{1/2}$ – leads to expressions for the path length, telescope length, and graze angle,

$$s = \sqrt{\Delta z^2 + (r_p - r_s)^2} + \frac{r_s}{r_p} f' + \frac{r_s r_p}{4f'}, \tag{2}$$

$$L = \Delta z + \frac{r_s}{r_p} f' - \frac{r_s r_p}{4f'}, \tag{3}$$

$$\tan 2\alpha = \frac{r_p - r_s}{\Delta z}. \tag{4}$$

These three equations must be satisfied for each shell, with $s - L$ and $f$ equal for all shells. Choosing $r_p$ as an independent variable, these three equations contain a total of four parameters that may vary by shell: $r_s, \Delta z, L$ and $\alpha$. There is one free parameter, allowing some design freedom, after matching the path length and focal length for all shells in a telescope. Rather than specifying $s - L$ directly, we define the geometry of one shell, called the *reference shell*, by defining any three of the four parameters for that shell. We choose to define $r_p^*, \Delta z^*$ and $\alpha^*$, where the asterisk superscript indicates parameters of the reference shell. The remaining three parameters $(s^*, L^*, r_s^*)$ can be calculated from Eqs. (2)-(4).

Once $s - L$ is determined by specifying the geometry of the reference shell, and one free parameter is eliminated by specifying one parameter or one relationship between parameters for each shell, Eqs. (2)-(4) can be solved symbolically and evaluated numerically. For example, if $L$ is specified for each shell (e.g., $L = L^*$), then Eqs. (2) and (3) can be solved for $r_s$ and $\Delta z$,

$$r_s = r_p \frac{2f'\left((s-L)(s+L) - r_p^2\right)}{4f'(f'(s-L) - r_p^2) + (s+L)r_p^2}, \tag{5}$$

$$\Delta z = L - \frac{2f'^2\left((s-L)(s+L) - r_p^2\right)}{4f'(f'(s-L) - r_p^2) + (s+L)r_p^2}\left(1 - \frac{r_p^2}{4f'^2}\right). \tag{6}$$

Since we require computation of $s - L$ with $\ll \lambda/14$ accuracy (Maréchal criterion), and Eqs. (5) and (6) contain products and sums of large (e.g., $f'$) and small (e.g., $s - L$) terms, numerical precision may be an important consideration in some cases. However, using Eqs. (2)-(6) for the example designs in this paper, double precision computation results in equal path lengths to less than 0.1 picometer, or $< \lambda/1000$.

### 2.3   Approximate design parameters

While an exact solution is necessary for accurate optical designs, approximate solutions provide intuition about the relationship between parameters. For $r_p \ll f \tan 2\alpha < f$ and $\alpha \ll 1$, we may approximate Eqs. (2) and (3) using a first-order Taylor series as,

$$s - L \approx \frac{1}{2} \Delta z \tan^2 2\alpha + \frac{r_p^2}{2f}, \tag{7}$$

$$f - L \approx \frac{f \Delta z \tan 2\alpha}{r_p}. \tag{8}$$

Typically, the path length calculated from Eq. (7) is microns different from that calculated from Eq. (2), but these equations are useful for understanding the trends of the geometric parameters.

Using Eqs. (4), (7) and (8), we analytically determine the four parameters for each shell with entrance radius $r_p$. There are four cases considered here, each where a different parameter is constant throughout the telescope. The resulting equations are shown in Table 1. To condense the table, we introduce the parameter $\chi = (r_{p,max}^2 - r_p^2)/(r_{p,max}^2 - r_p^{*2})$, where $r_{p,max} = \sqrt{r_p^{*2} + f \Delta z^* \tan^2 2\alpha^*}$ is approximately the maximum radius at which the path length of a shell can be matched to that of the reference shell (the denominators of Eqs. (5) and (6) are nearly zero when $r_p = r_{p,max}$). $\chi$ is a measure of how close $r_p$ is to $r_{p,max}$, and when $\chi \leq 0$, it is no longer possible to match the path length to that of the reference shell.

Table 1. Approximate expressions for the four geometric parameters, for four different cases. Here $\chi = (r_{p,max}^2 - r_p^2)/(r_{p,max}^2 - r_p^{*2})$.

| Quantity | Case 1<br>$L = L^*$ | Case 2<br>$\alpha = \alpha^*$ | Case 3<br>$\Delta z = \Delta z^*$ | Case 4<br>$r_p - r_s = r_p^* - r_s^*$ |
|---|---|---|---|---|
| $\dfrac{f - L}{f - L^*} =$ | 1 | $\dfrac{r_p^*}{r_p} \chi$ | $\dfrac{r_p^*}{r_p} \chi^{1/2}$ | $\dfrac{r_p^*}{r_p}$ |
| $\dfrac{\alpha}{\alpha^*} =$ | $\dfrac{r_p^*}{r_p} \chi$ | 1 | $\chi^{1/2}$ | $\chi$ |
| $\dfrac{\Delta z}{\Delta z^*} =$ | $\left(\dfrac{r_p}{r_p^*}\right)^2 \chi^{-1}$ | $\chi$ | 1 | $\chi^{-1}$ |
| $\dfrac{r_p - r_s}{r_p^* - r_s^*} =$ | $\dfrac{r_p}{r_p^*}$ | $\chi$ | $\chi^{1/2}$ | 1 |

Cases 1 and 2 are illustrated in Fig. 3, which shows several possible positions of the leading edges of the primary and secondary mirrors that change the path length while keeping $L$ or $\alpha$ constant. The dashed lines in Fig. 3 are defined by $r_p$ and $\alpha_f$, and $\alpha_f$ is related to $f$ by Eq. (1), so once $r_p$ and $f$ are chosen, the path length can only be adjusted by sliding the primary and secondary mirror positions along them (i.e., modifying ray segment B in Fig. 2). As discussed in Section 2.2, to match the path lengths for multiple shells, a delay must be added to a line going from the primary mirror to the focus, and the delay must get smaller as the shell radius increases. For Case 1 (where $L$ is fixed), the secondary mirror position must move along the sloped dashed line in Fig. 3a, which changes the radial gap $r_p - r_s$, the axial gap $\Delta z$, and the graze angle $\alpha$. For larger-diameter shells, the path must get straighter to reduce the delay.

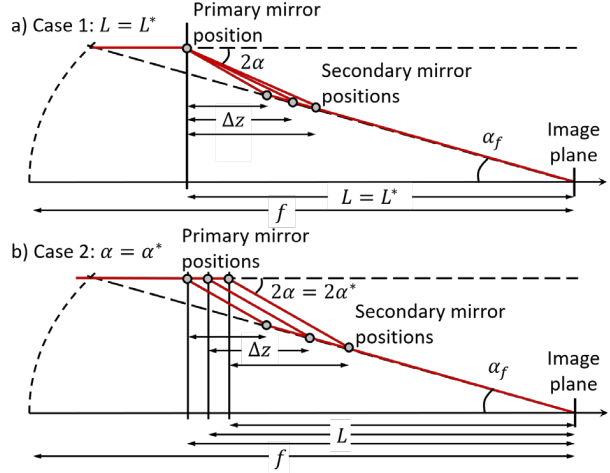

Figure 3. Illustration of varying the path length of a mirror shell while maintaining constant telescope length (a) or graze angle (b). The circles indicate possible positions of the leading edge of the primary and secondary mirrors.

While Case 1 allows a constant telescope length $L$, the graze angle $\alpha$ and axial gap $\Delta z$ must vary as a function of radius. Varying the graze angle introduces chromatic aberration, which we discuss in Section 4. Varying the axial gap is problematic if $\Delta z$ becomes large enough that it would be difficult to maintain sufficient alignment stability between the primary and secondary mirrors. Since for Case 1, $\Delta z$ varies roughly quadratically with shell radius, this effectively limits the range of shell radii that a diffraction-limited X-ray telescope can practically contain.

For Case 2 (where $\alpha$ is fixed, Fig. 3b), the primary and secondary mirrors are moved along the axial direction together such that the path between them remains at the same angle. The distance from the primary mirror to the focal plane, $L$, changes. For a large focal ratio $f/r_p \sim 10^5$, the two dashed lines in Fig. 3b are nearly parallel, and the required change in $L$ to effect a small change in path length can be very large. The change in length is $L - L^* \approx \frac{f}{r_p^*} \Delta z^* \tan 2\alpha^* \left(1 - \frac{f-L}{f-L^*}\right)$. If the focal ratio is large and if the quantity $(f-L)/(f-L^*)$ varies with radius (as in Cases 2-4), then the change in length can be quite large, often hundreds of meters. Cases 2-4 therefore seem impractical for a telescope with extremely long focal length.

There are other approaches we could take with respect to the free parameter. One example that may enable a large range of shell radii would be to specify two (or more) values of $L$, one for a set of large-radius shells, and another for a set of small-radius shells. The two sets of shells would be confocal and have the same path length, but would be axially displaced by many meters. However, the range of graze angles and axial gaps could be kept smaller in both sets than if they had the same $L$.

## 2.4   Mirror geometry for mas and μas telescope designs

Here we use Eqs. (4)-(6) to determine the geometry that matches path lengths for multiple shells in two telescope designs: a *μas telescope* design with shell diameters ranging from 2 m to 5 m, and a *mas telescope* design with shell diameters ranging from 220 mm to 600 mm. For these designs, we prescribe $L = L^*$ (Case 1 in Table 1). The innermost shell is chosen as the reference shell, and we add shells outward one at a time, choosing $\Delta r_p = h_m + L_m \tan \alpha_p$ (where $h_m$ and $L_m$ are the thickness and length of the mirrors, respectively) to avoid blocking on-axis X-rays as shown in Fig. 4. The geometric parameters of these designs are summarized in Table 2, and both mirror layouts are shown in Fig. 5. In this section, we consider only the geometry of these telescope designs, and evaluate various aspects of the optical performance in Sections 3-5.

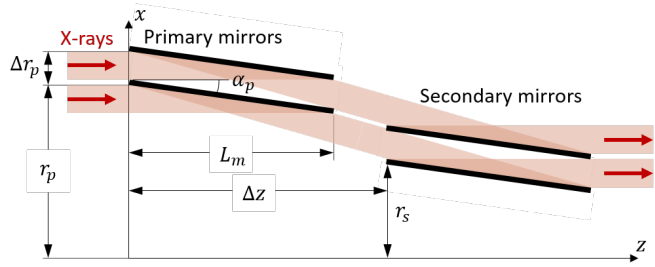

Figure 4. Nesting of shells in a Type 2 telescope.

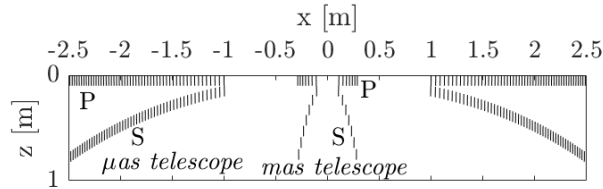

Figure 5. Cross-section of the mirror assembly for the *μas telescope* and *mas telescope* designs, approximately to scale. For clarity, only every 20th mirror is shown. X-rays enter from the top, reflect off of the inside of primary mirrors (P), off of the outside of secondary mirrors (S), then to the focal plane (300 km away for *μas*, 25 km away for *mas telescope* designs).

We limited the diameter of the outer-most shell to 5 m, a limit that would depend on future launch vehicles. The radii of the outermost and innermost shells affect both angular resolution and effective area. The outermost radius primarily affects the width of the central lobe of the PSF, while the innermost radius affects the fractional power in that central lobe. It is desirable to have a large difference in radius between the inner- and outer-most shells to maximize power in the core of the PSF, but this also significantly affects the axial gap $\Delta z$ and the range of graze angles in the telescope. The axial gap follows a quadratic relationship with radius, so if the axial gap is too large, alignment stability between the primary and secondary mirrors becomes more difficult. In these designs, we limit the end-to-end distance of the mirrors to 1 meter. We set the mirror length and thickness to 100 mm and 0.5 mm, respectively, matching the dimensions of the Lynx Reference Mission design [37]. When combined with our choice of $\Delta z^* = 100$ mm in both designs, this limits the range of radii to $r_p/r_p^* < 3$. To estimate the mirror mass, the mirror substrates are assumed to be silicon. Using longer mirrors may be possible but extends the mirror assembly along the optical axis, roughly proportional to the mirror length.

We calculated the prescription of the primary and secondary mirrors for $N = 0$ (a PH telescope). Calculating the prescriptions for $N = 1$ (a WS telescope) is difficult due to numerical instability, and we did not do this. The main benefit of a WS design is that aberrations such as coma are reduced compared to a PH design. However, we show (see Section 5) that for telescopes with large focal ratios, field curvature is the only important aberration for a PH design, and since the WS design generally has similar field curvature, we do not expect significant benefit from the WS design.

The graze angles for the primary and secondary mirrors of a Type 2 telescope are different by $\Delta \alpha = r_p/2f$, which is less than 1.3 arcsec for both telescope designs. The primary mirror is concave and the secondary mirror is convex, and both mirror surfaces slightly deviate from a perfect cone. The radial departure of the mirror surface from a cone is called the axial sag, and is nearly the same for each mirror. For the *mas telescope*, the difference in axial sag between the two mirrors ranges from 30 nm RMS at the inner-most shell to 2 nm RMS at the outer-most shell. For *μas telescope*, the values are about 100x smaller. The peak-to-valley axial sag is reported in Table 2.

**Table 2. Geometric parameters of two telescope designs. Number ranges span from innermost (reference) to outermost shells**

| Parameter | *μas telescope* | *mas telescope* |
|---|---|---|
| Focal length, $f$ [km] | 300 | 25 |
| Physical length, $L$ [km] | 299.0 | 24.21 |
| Shell radius, $r_p$ [m] | 1* – 2.5 | 0.11* – 0.30 |
| Graze angle, $\alpha_0$ [deg.] | 1.0* – 0.34 | 1.0* – 0.36 |
| Axial gap, $\Delta z$ [mm] | 100* – 731 | 100* – 766 |
| Radial gap, $r_p - r_s$ [mm] | 3.5* – 8.7 | 3.5* – 9.5 |
| Mirror length [mm] | 100 | 100 |
| Mirror thickness [mm] | 0.5 | 0.5 |
| Number of shells | 1033 | 131 |
| Mirror axial sag [nm PV] | 382* – 18 | 3558* – 164 |
| Total surface area [m$^2$] | 2400 | 35 |
| Total mirror mass [kg] | 2800 | 42 |

*Parameter of the reference shell

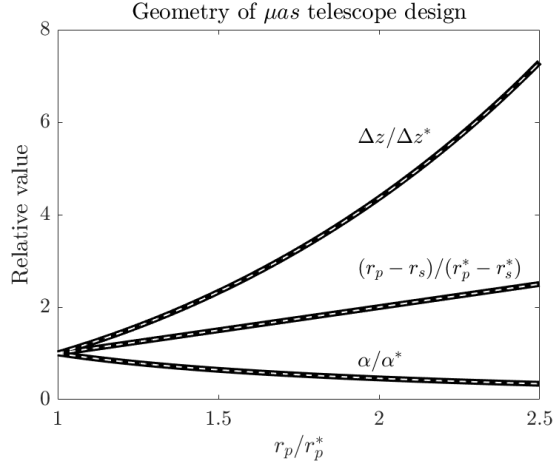

Figure 6. Relative values of the geometric parameters for the *μas telescope* design. The approximations (Case 1 of Table 1) are shown as dashed white lines.

One notable feature of a Type 2 telescope design is that the graze angle of the reference shell may be chosen to suit the science requirements, often without significantly affecting the telescope geometry. The graze angle impacts angular resolution, effective area and background noise, chromatic aberration, and total mirror surface area and mass. We set the graze angle of the reference shell to $\alpha^* = 1°$ in both designs as a balance between the competing attributes. Note that the graze angle decreases with increasing shell radius, unlike ordinary Type 1 x-ray telescopes.

The three geometric parameters (after $L = L^*$ is chosen) for the *μas telescope* design are shown in Fig. 6 as a function of $r_p/r_p^*$, using both the exact (Eqs. (4)-(6)) and approximate (Table 1) formulae. These relationships are nearly identical for the *mas telescope* design. The approximations (shown as white dashed lines) result in less than 0.1% error in the three geometric parameters for both telescope designs.

## 2.5 Extension to lower-resolution telescopes

It may be desirable to design lower-resolution telescopes than those presented here, because such telescopes might be easier to build and fly in the short-term. A lower-resolution telescope (with say 10 mas angular resolution) could have a significantly shorter focal length and looser pointing knowledge requirements. However, directly applying our design process to lower-resolution telescopes leads to several problems. For example, as the radius of the outer-most shell decreases, the axial length of the mirror assembly remains nearly fixed (see Fig. 5) unless the mirrors are made shorter. The field curvature also increases with decreasing telescope diameter (see Section 5), so the field of view over which the image is sharp would become smaller for a lower-resolution telescope.

Alternatively, we may use Eqs. (4)-(6) to design diffraction-limited mirror modules that consist of azimuthal segments from multiple shells (so each mirror module would have a roughly trapezoidal aperture). Many modules could be combined incoherently (i.e., the path length is not necessarily the same for each module) to obtain the desired effective area. The PSF of the telescope is then the sum of the PSFs from each module, and the resolution will depend on the size of the modules. Since the path lengths only need to be matched within a module, this approach may be simpler to build than a fully-coherent telescope. There are several potential complications to this approach, since the focal ratio could be significantly smaller and the field curvature may be different for different modules. We leave this as future work.

## 3. Resolution and effective area

### 3.1 Point spread function

Once the geometry of a telescope is defined, we can calculate the PSF and effective area that result from all of the shells (summarized in Table 3). The intensity distribution at the focal plane of a perfect telescope is the far-field diffraction pattern of the pupil function, and the effective collecting area may be calculated from this intensity distribution. For a diffraction-limited nested-shell grazing-incidence telescope, the pupil function is a set of annular rings, with phase shifts arising from chromatic aberration. Phase errors that vary within a shell and between shells will also be present, for example due to misalignment or surface errors, and the aperture function will also include blockage from support structures. For brevity, we ignore these in the present work, but they are important and will be considered in the future. Tschunko [22] first studied the point spread function arising from thin annular apertures, and Harvey [38] considered the incoherent sum of multiple annular apertures for X-ray telescopes. In contrast, here we consider the coherent sum of multiple annular apertures, since the path lengths are matched for each shell.

Table 3. Performance estimates of two telescope designs.

|  |  | μas telescope | mas telescope |
|---|---|---|---|
| Resolution[a] | | | |
| | 1 keV | 66 μas | 525 μas |
| | 5 keV | 14 μas | 111 μas |
| Effective area[b] | | | |
| | 1 keV | 4.3 m² | 645 cm² |
| | 5 keV | 2.9 m² | 440 cm² |
| Field of view (FOV)[c] | | 1.0 arcsec | 1.2 arcsec |

[a] Half-power diameter (HPD)

[b] Includes 20% loss from obscuration and other sources.

[c] See Section 5 for discussion of FOV.

The pupil function $G$ of the telescope, as a function of radial coordinate $x$ and energy $E$, is

$$G(x,E) = \sum_{m=1}^{M} \big(\rho(E,\alpha_m)\big)^2 \left[\mathrm{circ}(x/r_m) - \mathrm{circ}(x/r'_m)\right], \qquad (9)$$

where $\mathrm{circ}(a) = 1$ for $a \leq 1$ and $0$ otherwise. There are $M$ total shells, and the radii $r_m$ and $r'_m$ are the inner and outer radii of the $m^{th}$ annulus. The reflection coefficient $\rho$ is the complex amplitude ratio of the reflected to incident electric fields, and is squared here because there are two reflections at nearly-constant graze angle within each shell. The reflection coefficient is found using the Fresnel equations for the transverse-electric (TE) and transverse-magnetic (TM) polarization components [39],

$$\rho_{TM} = \frac{n^2 \sin\alpha - \sqrt{n^2 - \cos^2\alpha}}{n^2 \sin\alpha + \sqrt{n^2 - \cos^2\alpha}},$$

$$\rho_{TE} = \frac{\sin\alpha - \sqrt{n^2 - \cos^2\alpha}}{\sin\alpha + \sqrt{n^2 - \cos^2\alpha}}, \qquad (10)$$

where $n = 1 - \delta - i\beta$ is the index of refraction and $\delta$ and $\beta$ are functions of photon energy [40]. In calculating the PSF, we assume unpolarized light, i.e., $\rho = (\rho_{TE} + \rho_{TM})/2$. Throughout this paper we assume an iridium surface with root mean-square (RMS) roughness of 0.5 nm, and sufficient thickness (>20 nm) to make the effects of the underlying surface negligible. To account for roughness, we multiply the reflection coefficient by the Nevot-Croce factor [41], $\exp(-(4\pi\sigma/\lambda)^2 n \sin\alpha_t \sin\alpha\,/2)$, where $\sigma$ is the RMS roughness and $\cos\alpha_t = (1/n)\cos\alpha$ from Snell's law.

If the aperture is illuminated by a plane wave with intensity $I_0$, the intensity $I$ at the image plane is the squared-magnitude of the Fourier transform of the pupil function,

$$\frac{I(\theta,E)}{I_0} = \left| \frac{1}{iz\theta} \sum_{m=1}^{N} \big(\rho(E,\alpha_m)\big)^2 \left[r_m J_1(kr_m\theta) - r'_m J_1(kr'_m\theta)\right] \right|^2, \qquad (11)$$

where $J_1$ is the Bessel function of the first kind and order 1, $k = 2\pi/\lambda$, and $\theta$ is the angular position on the image plane. We have excluded some constant phase terms as they are the same for all shells and have no effect on the intensity.

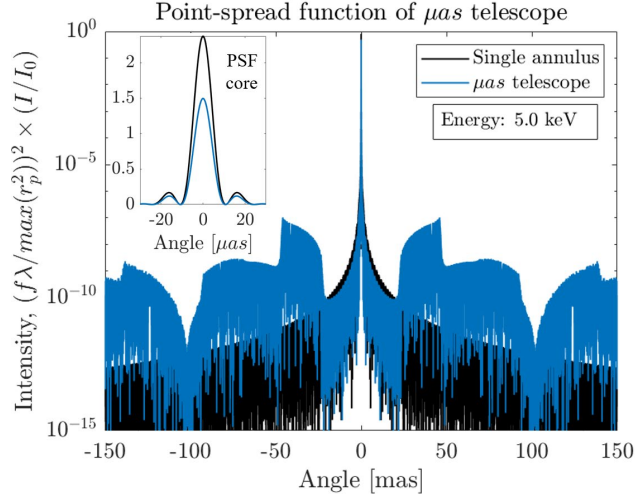

Figure 7. The PSF for the *µas telescope* design (blue) and a single annular aperture (black). Many diffraction orders from the nested shells are present, effectively representing background noise. The inset shows the central 60 µas of the PSF on a linear intensity scale.

The reflection coefficient is complex, leading to a phase shift from each reflection that depends on photon energy and graze angle. This chromatic aberration, and a strategy for its mitigation, is discussed in detail in Section 4. Ignoring this issue for now, and replacing $\rho(E, \alpha_m) = |\rho(E, \alpha_m)|$ in Eq. (11), we calculate and show the PSF for the *µas telescope* in Fig. 7. The PSF is compared to that of a *single-annulus* telescope, which is a telescope with one mirror shell that has 1 m inner diameter and 2.5 m outer diameter. Such a single-annulus telescope would be very long and probably impractical, but serves as a useful comparison.

The data in Fig. 7 show that a significant portion of the energy in the PSF of the *µas telescope* is far outside the central core. Several diffraction orders are clearly seen, resulting from the quasi-periodic radial spacing of the mirror shells. The inset of Fig. 7 shows the core of the PSF, showing there is minimal difference between the *µas telescope* and a *single annulus*, other than the fact that the intensity of the *µas telescope* PSF core is lower because some power is diffracted to larger angles. The PSF of the *mas telescope* exhibits similar behavior, except the angular values in the PSF core are increased by about a factor of 8.

### 3.2 Angular resolution

Angular resolution has several commonly-used definitions. The Rayleigh criterion defines the angular resolution as the angular radius of the PSF's first zero, obtaining $\theta_{Rayleigh} = 1.22\,\lambda/D$. However, by this definition, the angular radius of the first zeros of the PSFs of the *µas telescope* and a *single annulus* are both nearly equal to $\theta_{Rayleigh}$ and this provides limited information. Another measure of resolution is the half-power diameter (HPD), which is the angular diameter at which the encircled energy fraction (EEF) is equal to 0.50.

The EEF for an axisymmetric PSF is

$$\text{EEF}(\theta, E) = \frac{\int_0^\theta I(E, \theta')/I_0\, \theta' d\theta'}{\int_0^{\theta_{max}} I(E, \theta')/I_0\, \theta' d\theta'}, \qquad (12)$$

where $\theta_{max}$ is the maximum off-axis angle over which the PSF is integrated. For determining angular resolution, we choose to ignore background power, so we use $\theta_{max} = 10\,\theta_{Rayleigh}$. The angular resolution (and effective area, in the next section) is not very sensitive to the choice of $\theta_{max}$, since there is very little optical power between the core and the first background ring (which is at about 2000 $\theta_{Rayleigh}$ in Fig. 7).

The HPD is shown in Fig. 8 as a function of photon energy for the *µas* and *mas telescope* designs. The effects of chromatic aberration are also included in this figure, and discussed in Section 4. The HPD of the telescope (blue line) begins departing slowly from that of a *single annulus* (black dashed line) around 5 keV (λ=0.25 nm) in both telescopes. This is because the largest graze angle in each telescope is 1°, which is near the critical angle $\alpha_c \approx \sqrt{2\delta}$ for 5 keV photons reflecting off of an iridium surface [39]. At this energy, the inner-most shells of the telescope begin contributing significantly less optical power to the PSF, and the effective inner diameter of the telescope increases. This causes the power in the central lobe of the PSF to decrease, and eventually the half-power diameter occurs at angles larger than the first zero of the PSF, causing a jump in the HPD between 6-7 keV. Increasing the graze angle decreases the energy at which the telescope HPD diverges from the single annulus telescope.

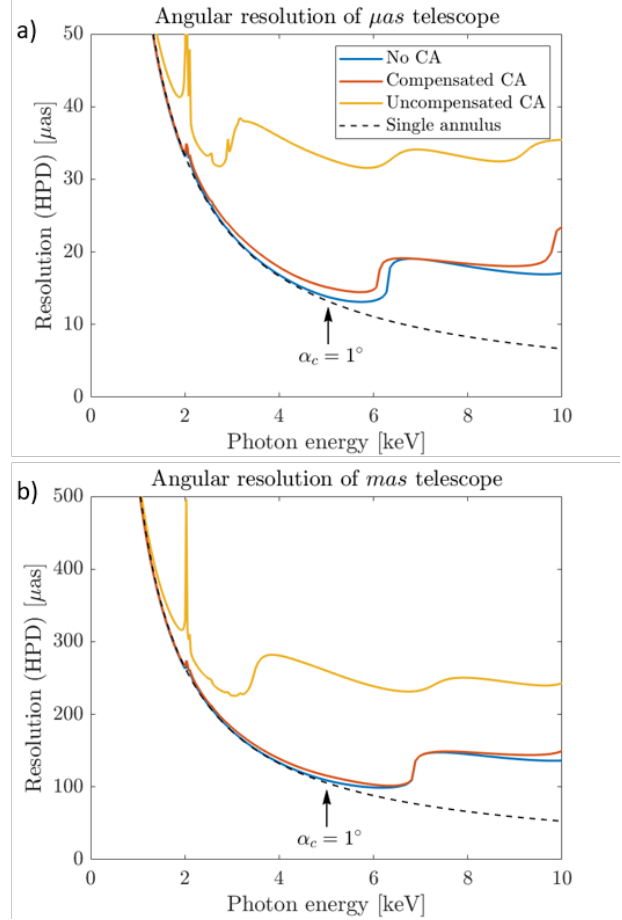

Figure 8. Half-power diameter of a) the *µas telescope*, and b) the *mas telescope*. The blue curve is the HPD accounting for all shells and assuming no phase error for each shell (an unrealistic case). The black dashed curve is for a *single annulus* telescope (see Section 3.1). The yellow curve shows the effects of chromatic aberration (CA), while the red curve shows the benefit of chromatic compensation (see Section 4). The spikes in the yellow curve between 2 and 3 keV correspond to absorption edges of iridium. The reference energy $E^*$ for chromatic aberration compensation is 7 keV in both cases. The arrow points to where the critical angle $\alpha_c = 1°$, which is the largest graze angle in the telescopes.

### 3.3 Effective area

The effective area of a diffraction-limited telescope is the integrated intensity in the core of the PSF (out to an angle $\theta_{max}$), given by

$$A_{DL}(E) = \int_0^{\theta_{max}} (I/I_0)\, f^2 \theta d\theta, \quad (13)$$

where $f$ is the focal length. For a non-diffraction-limited telescope, the effective area can be calculated by summing the product of the projected area of each shell and its squared-reflectivity (where reflectivity is $|\rho|^2$),

$$A_{NDL}(E) = \sum_{m=1}^{N} \pi(r_m^2 - r_m'^2)|\rho(E, \alpha_m)|^4. \tag{14}$$

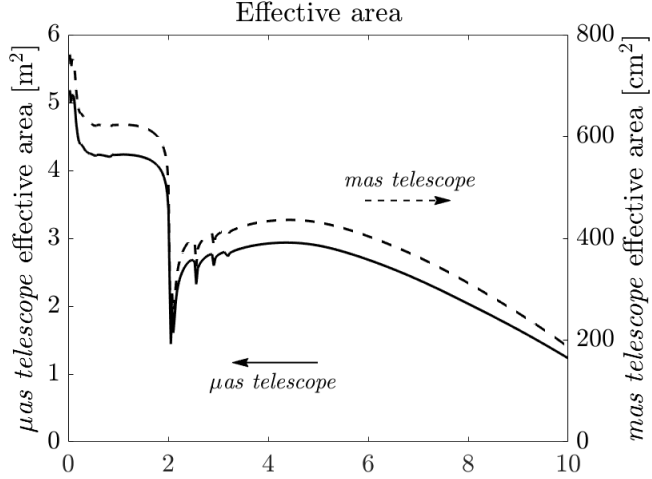

Figure 9. Effective area of the *μas telescope* (left) and the *mas telescope* (right).

The effective area for a diffraction-limited telescope is smaller than for a non-diffraction-limited telescope, because the quasi-periodic blockage from the mirror edges diffracts power away from the core of the PSF, as illustrated in Fig. 7. Since the mirror spacing is non-uniform, the intensity far from the PSF is many orders of magnitude smaller than in the core, and can be regarded as background power. For a one-dimensional uniform-period binary amplitude grating with open-area fraction $\eta$, the diffraction efficiency of the $0^{th}$-order is $\eta^2$ if the illumination is spatially-coherent, and $\eta$ if the illumination is spatially-incoherent. A diffraction-limited telescope is akin to the spatially-coherent case, and a non-diffraction-limited telescope is akin to the spatially-incoherent case. Therefore, $A_{DL}(E) \approx \eta A_{NDL}(E)$. This suggests that for a diffraction-limited telescope, blockage from mirror edges and structural supports has a much more significant impact on effective area than it does for a non-diffraction-limited telescope. Indeed, this simple analysis roughly agrees with the PSF shown in Fig. 7, where the open-area fraction of the aperture is 64%, and 67% of the power lies within the core of the PSF (i.e., encircled energy at $\theta = 0.13\ mas, E = 5\ keV$ and $\theta_{max} = 150\ mas$.) Similar agreement is found for the *mas telescope* design.

The effective area (Fig. 9) is calculated using Eq. (13), using $\theta_{max} = 10\ \theta_{Rayleigh}$, as we did when calculating angular resolution. If we use a larger value of $\theta_{max}$, then the calculated effective area is larger, but since the optical power that lies outside of the PSF core has very low intensity, it is indistinguishable from background noise and would not provide useful information. In calculating the effective area, we also assume an additional 20% loss from support structures, a value somewhat higher than assumed in the Lynx Reference Mission design [42]. Choosing a smaller graze angle for the reference shell can increase the effective area at high energies, but also causes more power to be diffracted to the wings of the PSF. For the *μas telescope* design, the mirror surface area is six times larger than the Lynx reference design mission, while the effective area is twice as large at 1 keV and 25 times larger at 6 keV. For the *mas telescope* design, the total mirror surface area is twice as large as Chandra and the effective area is similar. Based on the two examples given in the introduction, an effective area comparable to that of Chandra, about 1000 cm² will be sufficient to achieve compelling science goals.

### 3.4 Mirror assembly tolerances

Diffraction-limited performance requires maintaining equal length optical paths throughout the telescope to a small fraction of the shortest photon wavelength (typically λ/14 RMS [43]). This is challenging for X-rays since the wavelength is small, but the small graze angles used in X-ray telescopes result in the most stringent surface and alignment requirements being only on the order of 1 nm. A detailed mirror figure and assembly tolerancing study was performed for flat mirrors as part of the MAXIM mission study [23], and we will present a detailed tolerancing study for curved mirrors in a future paper. The tightest tolerances arise due to changes in radial distance between the surfaces of the primary and secondary mirrors. The error in path length $e_s$ resulting from errors in the radial gap $e_x$ and axial gap $e_z$ between mirrors is given by

$$e_s \approx 2e_x \sin\alpha + 2e_z \sin^2\alpha. \tag{15}$$

Since both the *µas* and *mas telescope* designs have mirrors with grazing angles varying between 0.36° and 1°, the tolerances are similar for the two designs. Maintaining path length errors below λ/14 RMS at 5 keV (λ=0.25 nm) requires maintaining radial gap errors below about 0.5-1.5 nm RMS and axial gap errors below about 30-250 nm RMS. The ratios of tolerances to nominal dimensions are 150 ppb for the radial gap, and 300 ppb for the axial gap, each roughly constant throughout both telescopes. Surface height errors and alignment errors both contribute to the path length error of Eq. (15).

It is expected that the primary and secondary mirrors shells would each be separated into small segments along the azimuthal direction. The alignment tolerances between a pair of primary and secondary mirror segments are generally significantly more stringent than the tolerances of pair-to-pair alignment [23]. The one exception is the pitch alignment of pairs to the telescope optical axis, where an angular pitch error $\Delta e_x/\Delta z$ leads to a path length error of

$$e_s \approx (r_p - r_s)\frac{\Delta e_x}{\Delta z}. \tag{16}$$

For both the *µas telescope* and *mas telescope* designs, $\Delta e_z$ must be maintained to 0.5-1.5 nm RMS (angular errors of 0.5-1 mas RMS) to obtain path length errors below λ/14 RMS.

New technology will be required to measure and fabricate mirror surfaces for diffraction-limited telescopes. While the Chandra Observatory has demonstrated (and the Lynx concept requires) 0.5 arcsec resolution, diffraction-limited performance requires axial profile error only several times more stringent than for those mirrors—certainly a challenge, but plausible in the coming decades. Diffraction-limited telescope mirror shells would require significantly smaller roundness errors than Chandra's or Lynx's mirrors. A more detailed analysis of curved mirror surface and alignment requirements will be presented in a future paper. Discussion of means to align and assemble mirrors to the required tolerances is beyond the scope of this paper, but clearly new technology will need to be developed. The most stringent tolerances lie in the sub-nanometer range—challenging to be sure—but examples of successful solutions in this domain can be found in the astrophysics and semiconductor manufacturing areas.

## 4. Chromatic aberration

An X-ray telescope that uses mirrors at multiple graze angles will have chromatic aberration. Chromatic aberration arises because reflection from X-ray mirrors by design occurs below the critical angle for total external reflection to achieve high reflectivity. In this regime, the reflection-induced phase shift varies with angle of incidence and photon energy. Above the critical angle, the reflection-induced phase shift approaches zero for all energies (since the index of refraction is smaller than 1 for most materials in the X-ray band) but the reflectivity is severely reduced. In this section, we discuss the effects of this chromatic aberration. We then describe how adjusting the physical path length and limiting the range of graze angles in the telescope can compensate for the effects of chromatic aberration sufficiently to enable diffraction-limited imaging over a wide energy band (0.1-10 keV in both the *µas* and *mas telescope* designs).

The reflection coefficient, defined in Eq. (10), is the ratio of reflected to incident wave amplitude. For X-rays, this coefficient is complex, and can be represented by a phase shift $\phi$ and magnitude $|\rho|$ as,

$$\rho(E, \alpha) = |\rho|e^{i\phi}. \tag{17}$$

Figure 10 shows the reflectivity $|\rho|^2$ and phase shift $\phi$ for a single reflection of unpolarized light from a smooth iridium surface as functions of graze angle and photon energy. A positive phase shift indicates that the reflected wave is advanced relative to the incident wave. Other common X-ray mirror coating materials (e.g., Au, Pt, W) exhibit similar behavior below the critical angle. In Section 3.1, we ignored the chromatic aberration by assuming $\rho = |\rho|$ in Eq. (11), but this assumption can be removed to evaluate the effect of chromatic aberration. The reflection-induced phase shift significantly degrades the angular resolution for higher energies, as illustrated by the yellow curve of Fig. 8 for the *µas* and *mas telescope* designs.

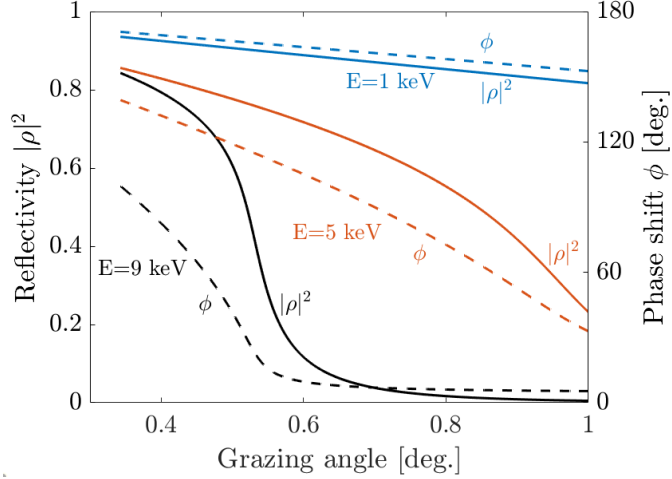

Figure 10. Reflectivity (solid lines) and reflection-induced phase shift $\phi$ (dashed lines) of a smooth iridium surface for three values of photon energy.

One approach to mitigating chromatic aberration is to adjust the physical path length, as a function of graze angle, to compensate for the phase shift at one energy (the reference energy, $E^*$) across the telescope aperture. The physical path lengths through the telescope are therefore unequal, introducing a wavelength-dependent phase error. The required path length adjustment for the reference energy is given by

$$\Delta s = \frac{hc}{E^*} \frac{\phi(E^*, \alpha)}{2\pi}, \qquad (18)$$

where $h$ is Planck's constant and $c$ is the speed of light. A positive value of $\Delta s$ will cause a phase delay that counters a positive phase shift $\phi$. Since $\phi$ ranges from 0 to $\pi$ (see Fig. 10), this requires a sub-nanometer path length adjustment, which can be incorporated by adjusting $s - L$ for each shell in Section 2.2. The path length change of Eq. (18) will result in a net phase shift $\phi'$,

$$\phi'(E^*, E, \alpha) = \phi(E, \alpha) - \frac{E}{E^*} \phi(E^*, \alpha). \qquad (19)$$

The red curve of Fig. 8 shows the HPD of the *μas* and *mas telescope* designs when chromatic aberration has been included but compensated by using Eq. (18). Using a reference energy $E^* = 7$ keV results in angular resolution that is very close to the resolution when chromatic aberration is not considered. The reference energy can be chosen depending on the science requirements. The Strehl ratio is defined as the peak intensity of the PSF of an aberrated optical system divided by that of the same system without aberrations, and a system with Strehl ratio over 0.8 is typically considered diffraction-limited. Comparing a telescope without chromatic aberration and one with compensated chromatic aberration, we find that the Strehl ratio is over 0.8 for the entire 0-10 keV energy band for both telescope designs, indicating that this compensation approach enables diffraction-limited performance.

To enhance reflectivity at some energies, compared to single-layer metal coatings, X-ray telescopes often employ multi-layer coatings that alternate low-electron-density materials with high-electron-density materials. We did not investigate chromatic aberration that would result from such coatings, but this would need to be considered in addition to the reflectivity when designing multi-layer coatings for a diffraction-limited telescope.

## 5. Detector size and off-axis aberrations

The field of view (FOV) of a diffraction-limited X-ray telescope is limited by detector size (or possibly vignetting). The maximum detector size over which a sharp image is formed is limited by off-axis aberrations and the position control accuracy of the detector relative to the optical axis of the telescope. Here we determine the position control tolerances, as a function of detector size, that enables sharp imaging over the entire detector.

The detector spacecraft, which is flying in formation with the mirror assembly spacecraft, may have position errors, so the line of sight corresponding to a pixel will vary over time and the detector position $(x_d, y_d, z_d)$ must be tracked to reconstruct images from photon counts. The line of sight, which is defined as the line connecting a particular pixel on the detector to the point where the principal surface intersects the optical axis of the mirror assembly (line $\overline{PQ}$ of Fig. 11), should be measured with angular accuracy better than the angular resolution, with respect to an inertial reference frame (pointing knowledge). Since X-ray photons are individually counted and time-tagged, control accuracy of the line of sight, which is accomplished by controlling the detector position relative to the optical axis, only needs to be smaller than the field of view (pointing control). Methods of measuring the angles of the optical axis relative to an inertial reference frame, and measuring the detector position relative to the nominal focus position, have been considered previously [17,29], but these important issues lie beyond the scope of this paper.

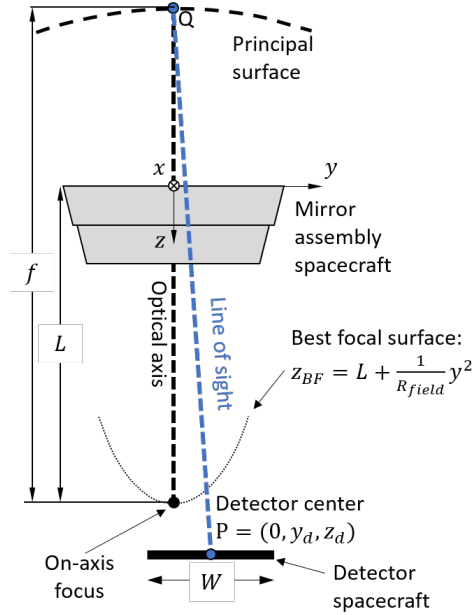

Figure 11. Depiction of mirror assembly spacecraft with principal surface, optical axis, and best focal surface illustrated. The detector spacecraft, with detector width $W$, is separate from the mirror assembly and located at position $P$. It must remain near the best focal surface $z_{BF}$, which is a paraboloidal surface with radius of curvature $R_{field}$ at the optical axis, to obtain a sharp image of an object along the line of sight $\overline{PQ}$.

The mirror assembly has off-axis aberrations and limited depth of focus, and together they limit how far the detector can deviate from the nominal focus position, and how large a flat detector can be used while maintaining a sharp focus over the entire detector. To estimate the position control tolerances for a particular detector size, we first discuss the telescope aberrations. The transverse ray aberration (TRA) analysis of Saha [44] can be simplified for a single shell of a long-focal length PH telescope ($N = 0$) with $f/D \sim 10^5$. We find that the most important aberration—curvature of field—is nearly the same for all shells.

Let us consider a ray at field angle $\Theta$, having a direction vector $(0, \sin\Theta, \cos\Theta)$, and intersecting the entrance aperture at location $(r\sin\gamma, r\cos\gamma, 0)$ given in polar coordinates. In the coordinate frame shown in Fig. 11, the nominal focus position is located at $(0, f\tan\Theta, L)$, and the ray intersects the detector plane at a position that deviates from the nominal focus position by $\Delta x$ and $\Delta y$. Saha [44] derives these TRA functions and expands them into functions of $\tan\Theta$, $\tan\Theta \sin 2\gamma$, and $\tan^2\Theta \sin\gamma$ for $\Delta x$ (and cosine terms for $\Delta y$). For a large focal ratio, only a few terms are significant, and the angular TRA functions, $\Delta\theta_x = \Delta x/f$ and $\Delta\theta_y = \Delta y/f$, simplify to

$$\Delta\theta_x \approx \frac{1}{4}\left(\frac{r}{f}\right)^2 \tan\Theta \sin 2\gamma - \left[\frac{R_1 + R_2 + r^2/2f}{R_2 R_1} r \tan^2\Theta + \frac{r}{f}\frac{(z_d - L)}{f}\right]\sin\gamma,$$

$$\Delta\theta_y \approx \frac{1}{4}\left(\frac{r}{f}\right)^2 \tan\Theta\, (2 + \cos 2\gamma)$$
$$- \left[\frac{R_1 + R_2 + r^2/2f}{R_2 R_1} r \tan^2\Theta + \frac{r}{f}\frac{(z_d - L)}{f}\right]\cos\gamma,$$

(20)

where $z_d$ is the axial position of the detector, and $R_1 = (2f(s-L) - r_p^2)/(s+L-2f)$ and $R_2 = (s^2 - L^2 - r_p^2)/(2f - 2s + r_p^2/2f)$ are the radii of curvature of the primary and secondary mirror surfaces at the optical axis, respectively [32].

The first term of Eq. (20) is due to coma, and the second term is due to curvature of field. Since $r/f \sim 10^{-5}$, the contribution to blur from the coma term, for a large field angle of 1° (a field angle that would cause many additional problems such as reflectivity loss) would still be less than 0.1 μas. For a WS telescope, it is exactly zero. Therefore, the only significant contribution to PSF blurring is the field curvature term. Using Eqs. (7) and (8) and the expressions for $R_1$ and $R_2$, the parabolic surface of best focus $z_{BF}$ is approximately,

$$z_{BF} \approx L - \frac{2\Delta z}{r_p^2} y^2 = L + \frac{1}{2R_{field}} y^2. \tag{21}$$

The best focal surface is axisymmetric, but in Eq. (21) it is independent of $x$ because of the chosen direction vector of the incoming ray. For the *μas telescope*, the radius of field curvature $R_{field}$ is approximately -2.5 m, and for the *mas telescope* it is only -30 mm. The field curvature is nearly constant for all shells in each telescope, since $\Delta z/r_p^2$ is approximately constant (see Table 1). We note that the telescope depth of focus, given by $DOF = \pm 2\lambda(f/D)^2$, is very large so even a flat detector can still achieve a large field of view with sharp focus.

We now turn to estimating the required position control tolerances for a flat detector of width $W$ (with $FOV = W/f$) and centered at position $(0, y_d, z_d)$, with normal vector parallel to the optical axis. The RMS angular blur due to curvature of field is the same in the x- and y-directions, and since $R_{field}$ is nearly constant for all shells in the telescope, we can approximate the RMS angular blur by integrating the square of the second term of Eq. (20) from $\min(r_p)$ to $\max(r_p)$. For simplicity we do not apply any weighting to the shells to account for different reflectivity or blockage in this estimate. The RMS angular blur at the edge of the detector $(0, y, z_d)$ is then,

$$\langle \Delta\theta \rangle \approx \left[ \frac{\Delta z}{r_p^2} y^2 + \frac{z_d - L}{2} \right] \frac{\max r_p}{f^2} \sqrt{1 + \left( \frac{\min r_p}{\max r_p} \right)^2}. \tag{22}$$

For $y \geq 0$, the largest RMS blur occurs at $(0, y_d + W/2, z_{BF} + \epsilon_z)$, where $z_d = z_{BF} + \epsilon_z$, and $\epsilon_z$ is the axial deviation of the detector from the best focal surface.

To provide a rough estimate of the required flight control tolerances for a given detector width, we assume that $\epsilon_z = y_d$ (i.e., the position control error in the z-direction is the same as that in the y-direction), and determine the position control error $y_d$ at which the RMS blur at the edge of the detector is equal to half of the HPD, i.e., $\langle \Delta\theta \rangle = \theta_{HPD}/2$. This occurs at,

$$y_d = \frac{\frac{f^2 \theta_{HPD}}{\max r_p}\left(1 + \left(\frac{\min r_p}{\max r_p}\right)^2\right)^{-1/2} - \frac{1}{2} W^2 \frac{\Delta z}{r_p^2}}{1 + 2W \frac{\Delta z}{r_p^2}}. \tag{23}$$

For the *mas telescope*, a 150 mm-wide detector composed of 15 μm pixels ($10^8$ total pixels, FOV 1.2 arc-seconds), and with a 125 μas HPD at 5 keV (λ=0.25 nm), the detector must be flown within ±300 mm of the best focal surface in the axial direction, and within 300 mm of the optical axis. Here the axial position tolerance is smaller than the depth of focus (DOF = ±0.9 m) because the detector's edge is farther from the strongly-curved best focal surface than is the center. For the *μas telescope*, even with an extremely large detector that is 1.5 m wide and with similar-sized pixels ($10^{10}$ total pixels, FOV 1.0 arc-second), and with a 14 μas HPD at 5 keV, the detector can be flown with ±1.5 m tolerance in the axial and lateral directions, which is similar to the depth of focus (DOF = ±1.8 m) due to the gently-curved best focal surface. As the detector size is reduced, the position control tolerances loosen.

Using a detector with smaller pixels has positive and negative effects on many aspects of the optical design, and choosing the pixel size for a particular design requires careful consideration. Smaller pixels will decrease the required focal length and make some aspects of formation flying easier. However, reducing the focal length of the telescope without modifying $r_p^*$, $\Delta z^*$ and $\alpha^*$ reduces $r_{p,max}$ (see Table 1), reducing the range of shell diameters that can be path-length-matched, which affects both the PSF and effective area. Reducing the focal ratio also decreases the DOF while field curvature is unaffected, and eventually the coma aberration may also become important. These issues could make formation flying tolerances more stringent. A smaller pixel size is not necessarily better, and the tradeoffs will require detailed analysis.

Counter-intuitively, the lateral position control tolerances (as measured in the mirror assembly frame of reference) can be larger than the detector size. In other words, the optical axis, which is fixed relative to the mirror assembly, may rotate relative to the line of sight without blurring the image. The mirror assembly does not need to remain fixed in inertial space to acquire a sharp image of an X-ray source, so position control-induced forces on this ultra-precision opto-mechanical system may be kept small. Of course, the image of the source must lie on the detector, so the line of sight must be controlled to a tolerance of some small fraction of the field of view, by controlling points P and Q in Fig. 11.

We list the fields of view calculated here in Table 3, since the detector position control tolerances are not overly tight for these detector sizes (and corresponding fields of view). If the detector can be tilted or curved, these tolerances may be relaxed somewhat, but we have not investigated this. Using a different optical prescription may also flatten the best focal surface while still achieving small effects of coma and astigmatism.

## 6. Conclusions

We have presented optical designs for two diffraction-limited nested-shell grazing-incidence X-ray telescopes, showing that such telescopes can have compact optical assemblies and achieve high angular resolution, large effective area, and wide field of view over a wide energy band. The *μas telescope* design, 5 meters in diameter, features 14 μas angular resolution and 2.9 m² effective area at 5 keV photon energy, enabling breakthrough scientific progress in a flagship-class mission. The *mas telescope* design, 0.6 meters in diameter, features 525 μas angular resolution and 645 cm² effective area at 1 keV. This telescope is significantly smaller, but could achieve sub-milli-arcsecond imaging, while serving as a pathfinder for the flagship mission. The mirror assembly for both telescopes is limited to less than 1 meter in length, which makes maintaining nanometer-position stability of mirrors much more feasible than for larger structures. We described X-ray telescope chromatic aberration which was successfully compensated by slightly adjusting the physical path length through each shell. We also investigated the effects of off-axis aberrations and found that curvature of field is the only significant aberration. Due to the large depth of focus, a flat detector can provide a sharp image over a very large field of view, around 1 arc-second in both telescopes. Furthermore, we found that lateral and axial position control tolerances of the detector spacecraft should be on the order of 1.5 m and 0.3 m for the *μas* and *mas telescope* designs, respectively. Although there are many technical challenges to achieving μas X-ray imaging, especially in making diffraction-limited mirror assemblies, we did not find any fundamental barriers. The design process we presented can be employed for designing future high-resolution X-ray telescopes targeted toward specific scientific goals. The topics we have presented could each be studied in far greater detail, but we hope this work has enumerated the major issues and provides a roadmap for others to follow.

## Funding

We acknowledge support from the National Aeronautics and Space Administration, Grants NNX16AD01G and NNX17AE47G.

## Acknowledgements

The authors wish to thank Paul Glenn, Timo Saha, Paul Schechter, David Windt, Youwei Yao, and William Zhang for their feedback.

## Disclosures

The authors declare no conflicts of interest.